\documentclass[page-classicl]{epl2} 

\usepackage[fleqn]{amsmath}
\usepackage{layout}
\usepackage{graphicx}
\usepackage[retainorgcmds]{IEEEtrantools}
\usepackage{cancel}
\usepackage{float}
\usepackage{url}
\usepackage{braket}
\usepackage{amssymb}
\usepackage{bigints}
\usepackage{amsthm}
\usepackage{mathtools}
\usepackage{cite}
\usepackage{textcomp}

\DeclarePairedDelimiter{\ceil}{\lceil}{\rceil}

\title{Softness, Polynomial Boundedness and Amplitudes' Positivity}

\author{Dong Bai \inst{1,2}\thanks{E-mail:\email{dbai@itp.ac.cn}\\Present address: School of Physics, Nanjing University, Nanjing, 210093, China.}}
\shortauthor{Dong Bai}

\institute{                    
  \inst{1} Key Laboratory of Theoretical Physics, Institute of Theoretical Physics, Chinese Academy of Sciences, Beijing 100190, China\\
  \inst{2} School of Physical Sciences, University of Chinese Academy of Sciences, No.19A Yuquan Road, Beijing 100049, China
}
\pacs{11.55.Bq}{}

\abstract{
In this note, we study the connections between infrared (IR) and ultraviolet (UV) behaviors of scattering amplitudes of massless channels by exploiting dispersion relations and positivity bounds. Given forward scattering amplitudes, which scale as $\mathcal{A}(s)\sim s^M$ in the IR ($s\to0$) and could be embedded into UV completions satisfying unitarity, analyticity, crossing symmetry and polynomial boundedness $|\mathcal{A}(s)|< c\, |s|^N$ ($|s|\to\infty$), with $M$ and $N$ integers, we show that the inequality $2\ceil*{\frac{N}{2}}\ge M \ge \ceil*{\frac{N}{2}}$ must hold, where $\ceil*{x}$ is the smallest integer greater than or equal to $x$. One immediate consequence of the above inequality is the bound on the UV growth of scattering amplitudes in terms of their IR behaviors. Our results could be useful in studies of massless higher spin particles, as well as the program of UV improvement and weakly-coupled UV completion.}

\begin{document}

\maketitle

It has been known for a long time that dispersion relations provide novel ultraviolet (UV)-infrared (IR) connections between low-energy effective field theories (EFTs) and their UV completions respecting various S-matrix axioms like unitarity, analyticity, crossing symmetry and polynomial boundedness \cite{Adams:2006sv}. These relations express low-energy forward scattering amplitudes (with identical in-states and out-states)\footnote{In this note, we presume the mathematical existence of forward scattering amplitudes. In particular, we presume that the $t\to0$ limit of scattering amplitudes is mathematically well-defined. {In other words, we are dealing with theories without massless exchanges for tree-level 2-to-2 scattering, as opposed to, e.g., general relativity.}\label{Footnote1}} in the deep IR as dispersive integrals of total cross sections along the positive $s$-axis extending into the deep UV. Celebrating implications of dispersion relations include positivity bounds on coefficients of certain higher derivative operators in EFTs, which have various applications in recent years \cite{Adams:2006sv,Distler:2006if,Manohar:2008tc,Low:2009di,Nicolis:2009qm,Komargodski:2011vj,Falkowski:2012vh,Urbano:2013aoa,Bellazzini:2014waa,Cheung:2014vva,Cheung:2014ega,Bellazzini:2015cra,Baumann:2015nta,Croon:2015fza,Hartman:2015lfa,Cheung:2016yqr,Komargodski:2016gci,Bellazzini:2016xrt,Bonifacio:2016wcb,deRham:2017avq,deRham:2017imi,deRham:2017zjm,Bellazzini:2017fep,deRham:2017xox}. In this note, we want to emphasize another connection between IR and UV behaviors of scattering amplitudes of \emph{massless} channels\footnote{By ``massless channels" we {mean} that in-states and out-states under consideration are all massless. In other words, it is gapless theories that are studied in this note. Certainly there could be massive particles in the deep UV.}, namely given forward scattering amplitudes, which scale as $\mathcal{A}(s)\sim s^M$ in the IR ($s\to0$) and could be embedded into UV completions satisfying unitarity, analyticity and crossing symmetry and polynomial boundedness $|\mathcal{A}(s)|< c\, |s|^N$ ($|s|\to\infty$), with $M$ and $N$ as integers, the inequality $2\ceil*{\frac{N}{2}}\ge M \ge \ceil*{\frac{N}{2}}$ must hold, where $\ceil*{x}$ is the smallest integer greater than or equal to $x$. This work is inspired by discussions in Ref.~\cite{Bellazzini:2016xrt} and many parts of this note could be viewed as variants or generalizations of that article. Although results in this note could have already been obtained in the 1960s, it is for the first time that they are displayed explicitly in literature as far as we know.

Dispersion relations, positivity bound, softness and polynomial boundedness of scattering amplitudes play important roles in our discussions, and we shall review them briefly as follows. 
Behind \underline{dispersion relations} are various S-matrix axiomatic properties including unitarity, analyticity, crossing symmetry and polynomial boundedness of scattering amplitudes. For \emph{massless} channels, they could be formulated as below:

(1) \emph{Unitarity}: S-matrix is unitary, i.e., $S^\dagger S = S S^\dagger = 1$, corresponding to the physical requirement of conservation of probability. As a result, for the forward scattering process $1^{h_1}_{a_1}2^{h_2}_{a_2}\to1^{h_1}_{a_1}2^{h_2}_{a_2}$, we have the optical theorem
\begin{equation}
\text{Im} \mathcal{A}^{h_1h_2}_{a_1a_2}(s+i\epsilon) = s\times\sigma(1^{h_1}_{a_1}2^{h_2}_{a_2} \to \text{Anything}; s),\qquad s\in \mathbb{R}^{+}.
\label{OpticalTheorem}
\end{equation}
Here and in the following, $\mathcal{A}$ denotes forward scattering amplitudes. $1^{h_1}_{a_1}$($2^{h_2}_{a_2}$) denotes massless particle 1(2) with four-momentum $k_1$($k_2$), helicity $h_1$($h_2$) and internal index $a_1$($a_2$). $s$ is the standard Mandelstam variable $s\equiv(k_1+k_2)^2$. The ``$+i\epsilon$" here is nothing but the usual Feynman prescription. Also, we have made use of the abbreviation $\mathcal{A}^{h_1h_2}_{a_1a_2}(s)\equiv\mathcal{A}(1^{h_1}_{a_1}2^{h_2}_{a_2}\to1^{h_1}_{a_1}2^{h_2}_{a_2};s)$.

(2) \emph{Analyticity}: The physical scattering amplitude is the real boundary value of an analytic function of complexified Mandelstam variables $s$, $t$ and $u$, with various simple poles and branch-cuts dictated by unitarity. For massless channels $\mathcal{A}^{h_1h_2}_{a_1a_2}(s)$ is at most as singular as simple pole at $s=0$. Furthermore, to derive dispersion relations we need to assume the Schwarz reflection principle
\begin{equation}
\mathcal{A}^{h_1h_2}_{a_1a_2}(s^*) =[\mathcal{A}^{h_1h_2}_{a_1a_2}(s)]^*,\qquad s\in \mathbb{C}.
\label{SchwarzReflection}
\end{equation}

(3) \emph{Crossing Symmetry}: Crossing symmetry of the forward scattering process $1^{h_1}_{a_1}2^{h_2}_{a_2}\to1^{h_1}_{a_1}2^{h_2}_{a_2}$ requires that
\begin{equation}
\mathcal{A}^{-h_1h_2}_{\bar{a}_1a_2}(s) =\mathcal{A}^{h_1h_2}_{a_1a_2}(-s),\qquad s\in \mathbb{C}.
\label{CrossingSymmetry}
\end{equation} 
Internal indices with a bar overhead label the states inside the complex conjugate representations carried by antiparticles.

(4) \emph{\underline{Polynomial Boundedness}}: Polynomial boundedness puts a stringent constraint on the UV behavior of complexified scattering amplitudes: 
\begin{equation}
|\mathcal{A}^{h_1h_2}_{a_1a_2}(s)|< c |s|^N\ {\text{or} \lim_{|s|\to\infty}|s|^{-N}|\mathcal{A}^{h_1h_2}_{a_1a_2}(s)|=0},\qquad\text{as}\ |s|\to\infty\text{, and }s\in \mathbb{C}.
\label{MasslessPolynomialBound}
\end{equation}
Here $N$ is some integer. Polynomial boundedness of this kind can be understood from the viewpoint of causality \cite{GellMann:1954db}. For gapped theories, we have $N\le2$ thanks to the famous Froissart bound $|\mathcal{A}^{h_1h_2}_{a_1a_2}(s)|\le\pi(s/m^2)[\log(s/s_0)]^2$ as $s\to\infty$ \cite{Froissart:1961ux,Martin:1962rt}. For gapless theories, the situation is a bit complicated, and there is no general result on what value $N$ should take (see, e.g., Ref.~\cite{Diez:2013vha} for a recent discussion). 

With the above properties dispersion relations could be derived easily. We start with the Laurent expansion of $\mathcal{A}^{h_1h_2}_{a_1a_2}(s)$ around $s=0$\footnote{{Rigorously speaking,} Laurent expansion cannot be done with respect to $s=0$ due to the presence of the branch-cuts $(-\infty,0)\cup(0,\infty)$. One has to first regularize the complex function $\mathcal{A}^{h_1h_2}_{a_1a_2}(s)$ to open the gap between the $s$-channel and $u$-channel branch-cuts, and recover the gapless scattering amplitude at the end of derivations. A suitable regularization scheme has to satisfy various requirements: $1^{\circ}$ it should indeed open the gap between the $s$-channel and $u$-channel branch-cuts; $2^{\circ}$ it should not change the imaginary part of $\mathcal{A}^{h_1h_2}_{a_1a_2}(s+i\epsilon)$ so that the optical theorem Eq.~\eqref{OpticalTheorem} holds for the regularized amplitudes on the gapped branch-cuts as well; $3^{\circ}$ it would be best if the regularization scheme introduces no extra unphysical simple poles; $4^{\circ}$ it should respect the Schwarz reflection principle Eq.~\eqref{SchwarzReflection}, crossing symmetry Eq.~\eqref{CrossingSymmetry} and polynomial bound Eq.~\eqref{MasslessPolynomialBound}. Take $s^2\log(-s^2)$ as an example. A suitable regularization could then be $s^2\log(-s^2+m^2)$, and one has the Laurent expansion $s^2\log(-s^2+m^2)=\log(m^2)s^2-\frac{s^4}{m^2}-\frac{s^6}{2m^4}+\cdots$ around $s=0$.  Although technical, these regularization schemes are, in fact, very important to surpass various obstructions associated with massless channels. It is useful to draw an analogy between the regularization schemes discussed here and dimensional regularization scheme which plays a fundamental role in proving the renormalizability of Yang-Mills theory. Comprehensive studies of possible realizations of such regularization schemes lie beyond the scope of this short note and are left for future studies. In this note, we shall simply assume the very existence of suitable regularizations for $\mathcal{A}^{h_1h_2}_{a_1a_2}(s)$. The following discussions are all worked out for regularized scattering amplitudes in the gapless limit, e.g., $m\to0$, although no special symbol is used to emphasize this point. Also, it is important to note that the regularization schemes introduced here are different from the common practice to add mass terms into gapless Lagrangians to turn the target theory into a gapped one. Although the latter practice also opens the gap between $s$-channel and $u$-channel branch-cuts, it often breaks crossing symmetry Eq.~\eqref{CrossingSymmetry} explicitly, introduces extra simple poles and leads to potential complications in higher spin theories.}
\begin{align}
{\mathcal{A}^{h_1h_2}_{a_1a_2}(s)  = \mathcal{A}^{h_1h_2}_{a_1a_2}(0) + s \times \mathcal{A}^{h_1h_2(1)}_{a_1a_2}(0) + s^2 \times \mathcal{A}^{h_1h_2(2)}_{a_1a_2}(0) + \cdots.}
\label{TaylorExpansion}
\end{align}
{Noticeably, there is no $1/s$ term in Eq.~\eqref{TaylorExpansion}. As mentioned in Footnote \ref{Footnote1}, we are considering theories without $t$ channel singularities (i.e., $1/t$ term). Then by crossing symmetry, these theories should also have no $1/s$ terms. There are also no higher negative powers of $s$ in Eq.~\eqref{TaylorExpansion}, such as the $1/s^2$ term, as terms of this kind would contradict with the axiom of analyticity.}

Introduce $L\equiv2\ceil*{\frac{N}{2}}$ which is an even integer. Then by Cauchy integral formula in complex analysis,
\begin{equation}
\mathcal{A}^{h_1h_2(L)}_{a_1a_2}(0) = \frac{1}{2\pi i}\oint_{\mathcal{C}}\frac{\mathrm{d}s}{s^{L+1}}\mathcal{A}^{h_1h_2}_{a_1a_2}(s).
\label{CauchyIntegral}
\end{equation}
The contour $\mathcal{C}$ is chosen to be the boundary of the cut complex plane $\mathbb{C}/[(-\infty,0)\cup(0,\infty)]$. Eq.~\eqref{CauchyIntegral} could be further simplified as
\begin{align}
{\mathcal{A}^{h_1h_2(L)}_{a_1a_2}(0) = \frac{1}{2\pi i}\left(\int_{-\infty}^0+\int_{0}^{\infty}\right)\frac{\mathrm{d}s}{s^{L+1}}\text{Disc}\,\mathcal{A}^{h_1h_2}_{a_1a_2}(s)+\mathcal{C}_{\infty}. } 
\label{CauchyIntegralSimplified}
\end{align}
 The integral $\mathcal{C}_{\infty}$ is done along the boundary contour at infinity. $\text{Disc}\,\mathcal{A}^{h_1h_2}_{a_1a_2}(s)$ is defined as
\begin{align}
\text{Disc}\,\mathcal{A}^{h_1h_2}_{a_1a_2}(s)\equiv\mathcal{A}^{h_1h_2}_{a_1a_2}(s+i\epsilon)-\mathcal{A}^{h_1h_2}_{a_1a_2}(s-i\epsilon)=2i\text{Im}\,\mathcal{A}^{h_1h_2}_{a_1a_2}(s+i\epsilon).
\label{Discontinuity}
\end{align} 
In the last step, we have used the Schwarz reflection principle Eq.~\eqref{SchwarzReflection}. By crossing symmetry Eq.~\eqref{CrossingSymmetry}, 
\begin{equation}
\text{Im}\,\mathcal{A}^{h_1h_2}_{a_1a_2}(-s+i\epsilon)=-\text{Im}\,\mathcal{A}^{-h_1h_2}_{\bar{a}_1a_2}(s+i\epsilon),\qquad s\in\mathbb{R}.
\label{CrossingSymmetryScatteringAmplitude}
\end{equation}
The boundary integral, on the other hand,  satisfies, 
\begin{equation}
\mathcal{C}_{\infty}\to0,
\label{ContourInfinity}
\end{equation}
thanks to the polynomial bound Eq.~\eqref{MasslessPolynomialBound} and the fact that $N\le L$.

Eq.~\eqref{CauchyIntegralSimplified}-\eqref{ContourInfinity} along with unitarity and the optical theorem Eq.~\eqref{OpticalTheorem} then give that
\begin{align}
\mathcal{A}^{h_1h_2(L)}_{a_1a_2}(0)&= \frac{1}{2\pi i}\left(\int_{-\infty}^0+\int_{0}^{\infty}\right)\frac{\mathrm{d}s}{s^{L+1}}\text{Disc}\,\mathcal{A}^{h_1h_2}_{a_1a_2}(s)\nonumber\\
&= \frac{1}{\pi}\int_{0}^{\infty}\frac{\mathrm{d}s}{s^{L}}\Big[\sigma(1^{h_1}_{a_1}2^{h_2}_{a_2} \to \text{Anything}; s)\nonumber\\
&+(-1)^L \sigma(1^{-h_1}_{\bar{a}_1}2^{h_2}_{a_2} \to \text{Anything}; s)\Big].
\end{align}
Taking into consideration that $L$ is an even integer, the above equation can be further simplified
\begin{align}
\mathcal{A}^{h_1h_2(L)}_{a_1a_2}(0)=&\frac{1}{\pi}\int_{0}^{\infty}\frac{\mathrm{d}s}{s^{L}}\Big[\sigma(1^{h_1}_{a_1}2^{h_2}_{a_2} \to \text{Anything}; s)+\sigma(1^{-h_1}_{\bar{a}_1}2^{h_2}_{a_2} \to \text{Anything}; s)\Big],
\label{DispersionRelation}
\end{align}
which is the $L$th subtracted dispersion relation. The $L$th subtracted \underline{positivity bound}
\begin{equation}
\mathcal{A}^{h_1h_2(L)}_{a_1a_2}(0)>0
\label{PositivityBound}
\end{equation}
is then simply followed from the fact that crossing sections $\sigma(1^{h_1}_{a_1}2^{h_2}_{a_2} \to \text{Anything}; s)$ are positive definite for interacting theories.

From Eq.~\eqref{DispersionRelation} it is straightforward to see how dispersion relations relate IR behaviors of scattering amplitudes to their UV behaviors. On the one hand, the right hand side (RHS) of Eq.~\eqref{DispersionRelation} involves a dispersive integral of total cross sections extending into arbitrary high energies, and the boundary integral $\mathcal{C}_{\infty}$ vanishes only because of the UV polynomial boundedness of scattering amplitudes. Generally, the RHS of Eq.~\eqref{DispersionRelation} is less tractable unless the full theory could be solved exactly from IR to UV. On the other hand, the left hand side (LHS) of Eq.~\eqref{DispersionRelation} concerns purely IR properties of scattering amplitudes, and could be calculated accurately by low-energy EFTs. It is the intractability of RHS and tractability of LHS of dispersion relations that motivate one to consider positivity bounds like Eq.~\eqref{PositivityBound}.

To characterize the IR behaviors of forward scattering amplitudes, it is also useful to introduce the notion of \underline{softness} which depicts how fast forward scattering amplitudes diminish as external momenta go to zero. For $\mathcal{A}^{h_1h_2}_{a_1a_2}(s)$, we have the Laurent expansion Eq.~\eqref{TaylorExpansion}. The key point here is that not all the terms of Eq.~\eqref{TaylorExpansion} are nonzero. For example, the forward scattering amplitude $\mathcal{A}(s)$ of the $P(X)$ theory\footnote{Here by $P(X)$ theory, we refer to theories of massless scalars whose Lagrangian could be parametrized as polynomials of $X=(\partial\phi)^2$. One such example could be $\mathcal{L}=\frac{1}{2}(\partial\phi)^2-\frac{\lambda}{4!}(\partial\phi)^4.$} only starts with the term $\sim s^2$, while terms before that are all zero. Generally,
\begin{equation}
\mathcal{A}^{h_1h_2}_{a_1a_2}(s) = s^M \times \mathcal{A}^{h_1h_2(M)}_{a_1a_2}(0) + s^{M+1} \times \mathcal{A}^{h_1h_2(M+1)}_{a_1a_2}(0) + s^{M+2} \times \mathcal{A}^{h_1h_2(M+2)}_{a_1a_2}(0)+\cdots,
\end{equation}
where the coefficient $\mathcal{A}^{h_1h_2(M)}_{a_1a_2}(0)$ is nonzero. The leading-order power $M$ is called softness of $\mathcal{A}^{h_1h_2}_{a_1a_2}(s)$, which could be symbolically denoted by $\mathcal{A}^{h_1h_2}_{a_1a_2}(s) \sim s^M$ as $s\to0$. Therefore, for $P(X)$ theory we have $\mathcal{A}(s)\sim s^2$ in the IR and the softness $M=2$. Also, it is noted by Ref.~\cite{Cheung:2014dqa,Cheung:2015ota} that the softer the amplitude is, the more symmetry there should be.

We are ready to prove the \underline{Main Result}:
\\[-1.5ex]

\emph{Given forward scattering amplitudes of massless channels, which scale as $\mathcal{A}^{h_1h_2}_{a_1a_2}(s)\sim s^M$ in the IR ($s\to0$) and could be embedded into UV completions satisfying unitarity, analyticity and crossing symmetry and polynomial boundedness $|\mathcal{A}^{h_1h_2}_{a_1a_2}(s)|< c\, |s|^N$ ($|s|\to\infty$), with $M$ and $N$ as integers, then the inequality $2\ceil*{\frac{N}{2}}\ge M \ge \ceil*{\frac{N}{2}}$ must hold, where $\ceil*{x}$ is the smallest integer greater than or equal to $x$.}

\begin{proof}
First notice that $M\ge0$, as $\mathcal{A}^{h_1h_2}_{a_1a_2}(s)$ is at most as singular as simple pole at $s=0$ and crossing symmetry requires the forward scattering amplitude to be an even function around $s=0$. Also, using the fact $\sigma\sim\frac{|\mathcal{M}|^2}{s}$ for the $2\to2$ scattering, we have that for massless particles the IR convergence of the dispersion relation in Eq.~\eqref{DispersionRelation} requires $M> L/2$ and since L is greater or equal than N (for the UV convergence) it follows that $M> N/2$. In fact, since M must be integer, it holds $M \ge \ceil*{\frac{N}{2}}$. 

{The rest proof goes as follows. The $L$th subtracted positivity bound requires that
\begin{equation}
\mathcal{A}^{h_1h_2(L)}_{a_1a_2}(s)\sim s^{M-L}>0,\qquad\text{as }s\to0,\nonumber
\end{equation}
which means that $M\le L\equiv2\ceil*{\frac{N}{2}}$.}
\end{proof}

One immediate consequence of the main result is the lower bound for the UV polynomial bound parameter $N$ in terms of the softness parameter $M$
\begin{equation}
{N> N_{\text{min}} \equiv 2\ceil*{\frac{M}{2}}-2.}
\label{NBound}
\end{equation} 
As mentioned before, the properties of polynomial boundedness of scattering amplitudes in gapless theories are less known compared to those in gapped theories. A common strategy to study polynomial bounds in gapless theories is to deform the original theories with mass terms to open the mass gap and turn the theories into gapped ones. This may be fine for theories involving only spin-0 and spin-$\frac{1}{2}$ particles, thanks to the fact that massless spin-0 and spin-$\frac{1}{2}$ particles have the same numbers of degrees of freedom as the massive ones, which makes it plausible to believe that no discontinuity comes into being when smoothly closing the mass gap to recover the gapless theories. If this is true, for these theories the UV polynomial bound parameter $N$ obeys $N\le2$, the same constraint inherited from gapped theories. Also, our constraint is less interesting for asymptotic free theories like Yang-Mills theory in which the UV behaviors of scattering amplitudes could be calculated explicitly. However, the situation is less clear for massless higher spin particles ($\text{spin}\ge2$), and it is wise to be open-minded for the possibility of $N>2$. Interesting theories of massless higher spin particles include the theory of interacting massless spin-2 particles proposed by Ref.~\cite{Wald:1986bj} which is ``normal" gauge invariant but not generally covariant (see also Ref.~\cite{Hinterbichler:2013eza,BaiXing:2016,Bai:2017dwf,Hertzberg:2016djj,Hertzberg:2017abn}), as well as various proposals of massless $\text{spin}>2$ particles (see e.g., Ref.~\cite{Boulanger:2006gr,Bekaert:2010hw}). {An unusual feature of these theories is that they all have ``too many" derivatives in the their interaction vertices. For instance, the massless spin-2 theory proposed by Ref.~\cite{Wald:1986bj} could have as many as eight derivatives in the quartic vertices (the cubic vertices could be tuned to be vanished, so there are no massless exchanges in 2-to-2 amplitudes), which leads to $\mathcal{A}(s)\sim s^4$ by naive dimensional analysis. Then by our bound, one has $N>2$ if this theory has any UV completions respecting various properties mentioned before.} 

 Also, our bound Eq.~\eqref{NBound} could be useful in the program of UV improvement and weakly-coupled UV completion \cite{Arkani-Hamed:2016,Arkani-Hamed:2017jhn}, where people try to modify target theories to make UV growth of scattering amplitudes as soft as possible. Take the nonlinear sigma model as an example, where one has $\mathcal{A}(s)\sim s^2$ in the IR and thus $M=2$. {Then according to Eq.~\eqref{NBound}, one has {$N>0$}. In other words, our bound tells that one could not modify nonlinear sigma model such that $|\mathcal{A}(s)|\sim |s|^{-\epsilon}<|s|^0$ $(\epsilon>0)$ in the UV while preserving various S-matrix axioms. When UV completing the nonlinear sigma model by a linear sigma model, one could see that the tree-level scattering amplitude $\mathcal{A}(s)\to \text{const}\sim|s|^0<|s|^{+\epsilon}$ in the UV, which is consistent with our bound.}  

In summary, inspired by discussions in Ref.~\cite{Bellazzini:2016xrt}, we work out a constraint on the UV behavior of forward scattering amplitudes of massless channels given the requirements that they admit meaningful UV completions. This constraint acts as a necessary condition for UV completibility and can be useful in studies of massless higher spin particles, as well as the program of UV improvement and weakly-coupled UV completion. 

\acknowledgments
{DB would like to thank Yu-Hang Xing and Lin Zhang for useful discussions. We would like to thank the anonymous referees for their patient reviewing and constructive suggestions.}

\end{document}